\documentclass[twocolumn,         
               showpacs,            
               preprintnumbers,     
               aps,                 
               prd,          	    
               a4paper,             
               superscriptaddress,      
               nofootinbib,         
               tightenlines,        
               floats,floatfix,11pt
               ]{revtex4-2}              
\usepackage{graphicx}  
\usepackage[margin=0.6in]{geometry}
\usepackage{dcolumn}   
\usepackage{bm}  
\usepackage[normalem]{ulem}
\usepackage{subcaption}
\usepackage[utf8]{inputenc}
\usepackage{amsmath,amssymb}
\usepackage{soul}
\usepackage{float}
\usepackage{makecell}
\usepackage[thinlines]{easytable}
\usepackage{array,booktabs}
\usepackage{lipsum} 
\usepackage[colorlinks=true,linkcolor=blue,citecolor=blue]{hyperref}
\usepackage{subcaption}
\usepackage{array,makecell}

\DeclareUnicodeCharacter{039B}{\ensuremath{\Lambda}}

\begin{document} 

\title{Quintessence scalar field and cosmological constant: Dynamics of a multi-component dark energy model }
\author{Prasanta Sahoo }
\email{prasantmath123@yahoo.com} 
\affiliation{Midnapore College (Autonomous), Midnapore, West Bengal, India, 721101}
\affiliation{NAS, Centre for Theoretical Physics \& Natural Philosophy, Mahidol University,
Nakhonsawan Campus, Phayuha Khiri, Nakhonsawan 60130, Thailand}
\author{Nandan Roy}
\email{nandan.roy@mahidol.ac.th (Corresponding Author)} 
\affiliation{NAS, Centre for Theoretical Physics \& Natural Philosophy, Mahidol University,
Nakhonsawan Campus, Phayuha Khiri, Nakhonsawan 60130, Thailand}
\author{Himadri Shekhar Mondal }
\email{himumath100@gmail.com} 
\affiliation{Midnapore College (Autonomous), Midnapore, West Bengal, India, 721101}


\begin{abstract}
This study explores the dynamics and phase-space behavior of a multi-component dark energy model, where the dark sector consists of a minimally coupled canonical scalar field and the cosmological constant, using a dynamical system analysis setup for various types of potential for which a general parameterization of the scalar field potentials has been considered. Several fixed points with different cosmological behaviors have been identified. A detailed stability analysis has been done and possible late-time attractors have been found. For this multi-component dark energy model, the late-time attractors are either fully dominated by the cosmological constant or represent a scenario where a combination of the scalar field and the cosmological constant dominates the universe. In this type of model, there is a possibility that the scalar field can become dynamical quite early compared to the standard era of dark energy domination. However, our analysis indicates that this early time contribution of the scalar field occurs deep in the matter-dominated era, not before the recombination era.
\end{abstract}

\maketitle

\section{Introduction}

Over two decades, various cosmological observations have confirmed the ongoing accelerated expansion of the universe \cite{SupernovaSearchTeam:1998fmf, SupernovaCosmologyProject:1998vns, Meszaros:2002np, Planck:2014loa, ahn2012ninth}, yet the cause of it remains unknown. In the standard model of cosmology, we consider the accelerated expansion to be driven by the cosmological constant \cite{padmanabhan2006dark}, which is consistent with current cosmological observations. However, it faces several challenges from both theoretical and observational perspectives, despite its significant achievements.

In addition to theoretical issues such as the cosmological constant problem and the fine-tuning problem, recent precision cosmological data have revealed a significant statistical discrepancy in the estimation of the Hubble parameter ($H_0$) between early-time and late-time observations. This discrepancy presents an additional challenge to the cosmological constant. Early universe measurements (e.g., CMB Planck \cite{Planck2020}, BAO \cite{BAO2017, BAO2011}, BBN \cite{BBN2021}, DES \cite{DES2018, DES:2018rjw, krause2017dark}) estimate $H_0 \approx (67.0 - 68.5)$ km/s/Mpc. In contrast, late-time measurements (e.g. SH0ES \cite{Sh0ES2019} and H0LiCOW \cite{Wong:2019kwg}) using time-delay cosmography find $H_0 = (74.03 \pm 1.42)$ km/s/Mpc. This $\simeq 5.3 \sigma$ discrepancy \cite{Riess_2022} hints towards new physics beyond $\Lambda$CDM in the dark energy sector.

Various dynamical dark-energy models have been suggested as alternatives for the cosmological constant \cite{amendola2010dark,Bamba:2012cp}. In dynamical dark energy models the equation of state of the dark energy changes over time \cite{copeland2006dynamics,Peebles2003,Armendariz2001,roy2022quintessence,Banerjee:2020xcn,Lee:2022cyh,Krishnan:2020vaf,Roy:2023uhc,Roy:2023vxk}. These models include but are not limited to quintessence, k-essence, and phantom-type scalar field models, where generally a scalar field is coupled with the matter minimally or non-minimally with a associated potential which can generate sufficient negative pressure to drive the accelerated expansion of the universe. Recent observations from DESI collaboration \cite{DESI:2024mwx, DESI:2024aqx} and other literature \cite{Roy:2022fif,Roy:2024kni,Dinda:2024kjf,Dinda:2024ktd} have pointed towards the evidence for the dynamical nature of the dark energy over the cosmological constant. 

The unknown nature of dark energy leads to the question of whether the dark sector consists of a single or multiple components. Recent literature \cite{LinaresCedeno:2021aqk, Roy:2023vxk,Sen:2021wld,Vazquez:2023kyx, Adil:2023ara} also shows that the multi-component nature of the dark energy sector might be more preferable by current observations than the cosmological constant. Current observations also suggest that the equation of state of the scalar field might have a phantom barrier $(w_\phi = -1)$ crossing in the recent past \cite{DESI:2024mwx,Roy:2024kni,Roy:2022fif}. In \cite{Cai:2009zp, Vazquez:2023kyx} it has been argued that there exists a no-go theorem which prohibits a dark energy model with a single degree of freedom to cross the phantom barrier. A multi-component model of dark energy is needed for the successful phantom barrier crossing.  In these types of multi-component models, the dark sector is considered to consist of a cosmological constant together with other dark energy components like a scalar field \cite{LinaresCedeno:2021aqk} or a fluid \cite{Sen:2021wld}, or to be composed of multi-scalar fields \cite{Roy:2023vxk}. In \cite{LinaresCedeno:2021aqk} a multi-component dark energy model has been proposed with a phantom scalar field and a cosmological constant, in \cite{Sen:2021wld} cosmological constant is considered with a fluid as dark energy and in \cite{Roy:2023vxk} a generic form of the quintom model has been considered with quintessence and a phantom field as the component of dark energy.  In addition, all these works suggest that multi-component models fit the data better than single component models and can significantly reduce Hubble tension compared to the $\Lambda$CDM model.

One can study the phenomenology of these multi-component dark energy models by comparing them against the state of the art cosmological observations. One can also use the powerful techniques of dynamical system analysis to investigate the phase-space behavior of these models without finding an exact solution. Dynamical system analysis is particularly useful when we are interested in the asymptotic behaviour of the system. Since these models have multiple components, it is crucial to understand their late-time behavior to identify which component might dominate the universe as the late-time attractor. In this paper, we use dynamical system analysis to investigate a multi-component dark energy model where the dark sector comprises a minimally coupled canonical scalar field, known as a quintessence field, and a cosmological constant. Our primary goal is to examine this composite model's phase space behaviour and dynamics using the dynamical systems approach. By applying appropriate variable transformations, we convert the system of equations into a set of autonomous equations. Subsequently, these equations are recast in polar form to facilitate mathematical handling. We have considered a parameterization of the scalar field potential that can incorporate various forms of potential to keep our analysis as general as possible since there is no consensus on the choice for the form of the scalar field potential. A detailed stability analysis of the system is presented. Additionally, we explore the dynamics of this model by numerically solving it, revealing the possibility of early time contribution of the scalar field.

The manuscript is structured as follows. In Section II, we cover the mathematical setup of the model. Section III forms the autonomous system, which is then transformed into polar form in Section IV. Section V provides a detailed stability analysis of the system.  In Section VI, we numerically investigated the evolution of the model for different classes of potentials. Finally, in Section VII, we summarise and conclude our findings.

\section{Mathematical Background}
In a spatially flat universe described by the standard FLRW metric which includes relativistic components, matter components, a minimally coupled canonical scalar field known as the quintessence field, and the cosmological constant; the Friedmann equations can be expressed as follows: 

\begin{equation}\label{rsh01}
 H^{2}=\frac{\kappa^{2}}{3}\left(\rho_{m}+\rho_{r}+\rho_{\phi}+\rho_{\Lambda}\right),\end{equation}

  \begin{equation}\label{rsh02}
\dot{H}=-\frac{\kappa^{2}}{2}\sum_i\left( p_{i}+\rho_{i}\right) .
\end{equation}


Here, $\kappa^{2}=8\pi G$, with $H=\dot{a}/a$ denoting the Hubble parameter and $a(t)$ representing the scale factor. The terms $p_{\phi}=\frac{1}{2}\dot{\phi}^{2}+V(\phi)$ and $\rho_{\phi}=\frac{1}{2}\dot{\phi}^{2}-V(\phi)$ correspond to the pressure and energy density of the scalar field. The subscripts $m$, $r$, $\phi$, and $\Lambda$ refer to matter, radiation, quintessence, and the cosmological constant, respectively. The pressure $p_i$ and the energy density $\rho_i$ for each species $i$, namely $m$, $r$, $\phi$, and $\Lambda$, are interrelated through the relation $p_i=w_i \rho_i$, where

\[w_{i}=\begin{cases}
\frac{1}{3}, & \text{for  relativistic  matter}\\
0, & \text{for  non-relativistic  matter} \\
-1, & \text{for } \Lambda .
\end{cases}
\]

The Klein-Gordon equation for the scalar field and the continuity equations, respectively, can be expressed as follows:

\begin{equation}\label{rsh03}
\ddot{\phi}+3H\dot{\phi}+\frac{dV(\phi)}{d\phi}=0 , 
\end{equation}

\begin{equation}\label{rsh04}
 \dot{\rho_{i}}=-3H(p_{i}+\rho_{i}),\quad \forall \quad i = m,r,\phi , \Lambda .
\end{equation}

The density parameter for a given species `i' is expressed as $\Omega_{i}=\frac{k^{2}\rho_{i}}{3H^{2}}$. Consequently, the Friedmann constraint can be written as,

\begin{equation}\label{rsh05}
 \Omega_{m}+\Omega_{r}+\Omega_{\phi}+\Omega_{\Lambda}=1.
\end{equation}

\section{The Dynamical System }

To understand the phase-space behavior of the system, one needs to introduce a new set of dimensionless variables to write it as an autonomous system. Here we consider the following set of dimensionless transformations:

\begin{align}\label{rsh06}
 x^{2} & = \frac{\kappa^{2}\dot{\phi}^{2}}{6H^{2}}, \quad y^{2}= \frac{\kappa^{2}V (\phi)}{3H^{2}}, \\ \nonumber
 \lambda &= -\frac{1}{k V}\frac{dV (\phi)}{d\phi}, \Gamma =\frac{V \frac{d^{2}V}{d\phi^{2}}}{\left( \frac{dV}{d\phi} \right)^{2}}.
\end{align}

With the help of the above transformations, the system can be reduced to a set of autonomous equations;

\begin{subequations}\label{cartesian_autonomous:1}
\begin{align}
    x^{\prime}&=-3 x+\sqrt{\frac{3}{2}} \lambda y^{2}+\frac{1}{2} x\left( 6x^{2}+3\Omega_{m}+4 \Omega_{r}\right),\label{eq:x1} \\
    y^{\prime}&=-\sqrt{\frac{3}{2}} \lambda x y+\frac{1}{2} y\left(6x^{2}+3\Omega_{m}+4 \Omega_{r}\right),\label{eq:y1}\\
     \Omega_m^{\prime}&= \Omega_m\left(-3+6x^{2}+3\Omega_{m}+4 \Omega_{r}\right), \\
 \Omega_r^{\prime}&= \Omega_r\left(-4+6x^{2}+3\Omega_{m}+4 \Omega_{r}\right),\\
 \lambda^{\prime}&=-\sqrt{6} x (\Gamma-1)  \lambda^{2}=-\sqrt{6} x f. \label{eq:lam1} 
\end{align}
\end{subequations}

Here, the derivatives are with respect to $N=ln\left( \frac{a}{a_{0}} \right)$ and $f = (\Gamma-1)  \lambda^{2}$, where $a_{0}$ is the present value of the scale factor. In such a scenario, the total Equation of State (EoS) can be written as

\begin{equation}\label{rsh07}
 \omega_{tot} \equiv \frac{p_{tot}}{\rho_{tot}}=-1+2x^{2}+\Omega_{m}+\frac{4}{3}\Omega_{r}.
\end{equation}

The system of equations given in Eq.$(\ref{cartesian_autonomous:1})$ cannot be a closed system due to the arbitrariness of the form of the scalar field potential $V(\phi)$. To close the system, one needs to find $f$ as a function of variables $x,y,\lambda$. Since by definition $f$ contains the derivative of the potential, the form of it will depend only on the choice of the potential. Since there is a lack of consensus on the form of the potential of the scalar field a wide variety of potentials has been used in the literature to study these models.

In one approach, a specific potential is selected, determining the corresponding form of $f$. This method has already been utilized in the literature; for examples, see  \cite{BURD1988929, 1988ApJ...325L..17P, Barrow_1995, Parsons_1995, PhysRevD.62.103517, GONG2006286, doi:10.1142/S0218271800000542, Lidsey_2002, Matos_2000, Ure_a_L_pez_2000, Matos_2009, Leyva_2009, Leon_2009, Aref_eva_2010, Escobar_2012, Escobar__2012, del_Campo_2013, article_2013, Fadragas_2014, leon2013quintomphasespaceexponentialpotential, Leon_2018, Leon_2023, Cid_2016, Paliathanasis_2015, Arapo_lu_2019, Bahamonde_2018, Quiros_2019, Leon_2019, Leon_2020, Leon__2020, P_rez_2021, Khyllep_2022, Bhanja_2023, halder2024interactingphantomdarkenergy, paliathanasis2024revisedarkmatterphantomscalar}. Conversely, one can start with a specific form of $f$ and derive the corresponding potential using the definitions of $y$, $\lambda$, and $\Gamma$ variables as given in Eq.(\ref{rsh06}).
To ensure our analysis applies to a wide class of potentials, we adopt the form $f = \alpha_{0} + \alpha_{1}\lambda + \alpha_{2}\lambda^{2}$, where $\alpha_{0}$, $\alpha_{1}$, and $\alpha_{2}$ are real-valued parameters. This particular parametrisation of $f$ was first proposed in \cite{Roy:2017uvr}.  Different choices of the $\alpha_{0}, \alpha_{1}$ and $\alpha_{2}$ parameters will correspond to different classes of potentials. In Table-\ref{tab:potentials}, we list eight distinct classes of potentials, each corresponding to different choices of the $\alpha$ parameters. The classification was made based on whether certain $\alpha$ parameters are zero or not.  This approach has been taken to facilitate our analysis for a wide class of potentials in a single setup. A similar approach, but for a different choice of dynamical variables, has also been used in \cite{Roy:2018nce}. Within the potentials listed in Table-\ref{tab:potentials}, $V_{3}$ and $V_{8}$ 
 has been already studied in \cite{Copeland_2024} and \cite{Ramadan:2023ivw} respectively for early time solutions. Whereas in \cite{Roy:2014yta} dynamical system analysis for $V_7$ has been studied for the quintessence field.\\

\begin{table*}[!hbt]
    \centering
    \scalebox{1.18}{
    \begin{tabular}{|c|c|c|}
       \hline {Label} & {Structure of $f$} &  {Potential $V(\phi)$} \\ \hline
         $V_{1}$ & $\alpha_{0}\neq 0$, $\alpha_{1}\neq 0$, $\alpha_{2}\neq 0$ & $A\hspace{0.02cm} exp \left[\frac{\alpha _{1} k \phi -2 \log \left(\cosh \left(\frac{1}{2} \sqrt{\alpha _{1}^{2}-4 \alpha _{0} \alpha _{2}} k (\phi +B)\right)\right)}{2 \alpha _{2}}\right]$   \\  \hline
        $V_{2}$ &  $\alpha_{0}= 0$, $\alpha_{1}\neq 0$, $\alpha_{2}\neq 0$ & $A\hspace{0.02cm} \left(e^{\alpha _{1} k \phi }\right){}^{\frac{1}{\alpha _{2}}} \left(\alpha _{1} \alpha _{2} k \left(e^{\alpha _{1} k \phi }+\alpha _{2} e^{\alpha _{1} A k}\right)\right){}^{-\frac{1}{\alpha _{2}}}$  \\  \hline
        $V_{3}$ &  $\alpha_{0}\neq 0$, $\alpha_{1}= 0$, $\alpha_{2}\neq 0$ &  $A\hspace{0.02cm} \cos ^{-\frac{1}{\alpha _{2}}}\left(\sqrt{\alpha _{0} \alpha _2} k (\phi +B)\right)$  \\  \hline
        $V_{4}$ &  $\alpha_{0}\neq 0$, $\alpha_{1}\neq 0$, $\alpha_{2}= 0$ &  $A\hspace{0.02cm} exp \left[ \frac{\alpha _0 k^2 \phi -c_1 e^{-k\alpha _{1}  \phi }}{\alpha _{1} k} \right]$  \\  \hline
         $V_{5}$ &  $\alpha_{0}\neq 0$, $\alpha_{1}= 0$, $\alpha_{2}= 0$ &   $A\hspace{0.02cm}  exp\left[\frac{1}{2} \alpha _{0} k^{2} \phi ^{2}+B \phi \right]$  \\  \hline
          $V_{6}$ &  $\alpha_{0}= 0$, $\alpha_{1}\neq 0$, $\alpha_{2}= 0$ &   $A \hspace{0.02cm}  exp\left[-\frac{B e^{-k\alpha _{1}\phi }}{\alpha _{1} k}\right]$  \\  \hline
           $V_{7}$ &  $\alpha_{0}= 0$, $\alpha_{1}= 0$, $\alpha_{2}\neq 0$ &   $A\hspace{0.02cm}  \left(\alpha _{2} \phi +B\right){}^{-\frac{1}{\alpha _{2}}}$  \\  \hline
        $V_{8}$ &  $\alpha_{0}= 0$, $\alpha_{1}= 0$, $\alpha_{2}= 0$ &   $A e^{B\phi}$  \\  \hline
         \end{tabular}
         }
    \caption{ A list of various classes of potentials based on the different choices for the $\alpha$ parameters. Here $A$ and $B$ are integration constants.}
    \label{tab:potentials}
\end{table*}

\section{REPRESENTATION OF THE 
 DYNAMICAL SYSTEM IN POLAR FORM}

The use of polar coordinates to study the cosmological system is often useful, as shown in \cite{Roy:2018nce,Roy:2017uvr,Roy:2013wqa} because the new variables are directly related to the cosmological parameters and the mathematical handling of the system becomes easier. To transform the system of equations Eq.$(\ref{cartesian_autonomous:1})$ into polar form, we use the transformation $x=r \cos \theta$ and $y=r  \sin \theta$, where $r^{2}=x^{2}+y^{2}=\Omega_\phi$. With the above choice, the system of equations reduces to

\begin{subequations}\label{polar_autonomous:1}
\begin{align}
    r^{\prime}&=\frac{r}{2}\left(-3 -3\cos (2\theta) +6r^{2}\cos^{2}\theta +3 \Omega_{m}+4\Omega_{r} \right),\label{eq:r2} \\
    \theta^{\prime}&=\frac{1}{2}\sin\theta \left(6\cos\theta -\sqrt{6}\lambda r \right),\label{eq:th2}\\
     \Omega_m^{\prime}&= \Omega_m\left(-3+6r^{2}\cos^{2}\theta +3 \Omega_{m}+4\Omega_{r} \right) ,\\
 \Omega_r^{\prime}&= \Omega_r\left(-4+6r^{2}\cos^{2}\theta +3 \Omega_{m}+4\Omega_{r} \right),\\
 \lambda^{\prime}&=-\sqrt{6} r \cos\theta \left( \alpha_{0}+\alpha_{1}\lambda +\alpha_{2}\lambda^{2} \right) .\label{eq:lam2} 
\end{align}
\end{subequations}

In addition, the Friedmann constraint given in Eq.$(\ref{rsh05})$ takes the following form

\begin{equation}\label{rsh08}
 r^{2}+\Omega_{m}+\Omega_{r}+\Omega_{\Lambda}=1.
\end{equation}

The total equation of state and the equation of state of the scalar field can be represented in the polar form as follows, 

\begin{equation}\label{rsh09}
 \omega_{tot} =-1+2r^{2}cos^{2}\theta +\Omega_{m}+\frac{4}{3}\Omega_{r},
\end{equation}

\begin{equation}\label{rsh10}
 \omega_{\phi} \equiv \frac{p_{\phi}}{\rho_{\phi}}=cos\hspace{0.02cm}2\theta .
\end{equation}

\begin{table*}[!h]
    \centering
    \scalebox{1.25}{
    \begin{tabular}{|c|c|c|c|c|c|c|}
       \hline {\vtop{\hbox{\strut Equilibrium}\hbox{\strut points}}} & { $r_{c}$} &{$\theta_{c}$} & {$\Omega_{m c}$} & {$\Omega_{r c}$} & {$\lambda_{c}$}&   {Potentials} \\ \hline
        $r$ &  $0$ & $\frac{n\pi}{2}$ & $0$ & $1$ & $\lambda$  &  $V_{1}- V_{8}$  \\  \hline
        $m$ & $0$  & $\frac{n\pi}{2}$ & $1$ & $0$ & $\lambda$ & $V_{1}- V_{8}$  \\  \hline
        $e1$ & $ 1$ &  $n\pi$ & $0$ & $0$ & $0$ & $V_{2}, V_{6}-V_{8}$ \\  \hline
         $e2$ & $ 1$& $n\pi \pm \frac{\pi}{2}$ & $0$ & $0$ & $0$  & $V_{1}- V_{8}$ \\  \hline
        $e3$ &  $ 1$ & $ n\pi$ & $0$ & $0$ & $A_{\pm}$  & $V_{1}- V_{3}, V_{7}$  \\  \hline
         $e4_{\pm}$ &  $ 1$ & $n\pi \pm cos^{-1}\left( 
        \frac{A_{\pm}}{\sqrt{6}} \right)$ & $0$ & $0$ & $A_{\pm}$ &  $V_{1}- V_{3}, V_{7}$  \\  \hline
         $c$ & $0$ &  $\frac{n\pi}{2}$ & $0$ & $0$ & $\lambda$  & $V_{1}- V_{8}$  \\  \hline
          $er1_{\pm}$ & $ 2E_{\pm}$  & $2n\pi \pm cos^{-1}\left( \sqrt{\frac{2}{3}}\right) $ & $0$ & $B_{\pm}$ & $A_{\mp}$ & $V_{1}, V_{3}$  \\  \hline
          $er2_{\pm}$ & $- 2E_{\pm}$  & $(2n+1)\pi \pm cos^{-1}\left(- \sqrt{\frac{2}{3}}\right) $ & $0$ & $B_{\pm}$ & $A_{\mp}$ & $V_{1}, V_{3}$  \\  \hline
          $er3_{\pm}$ & $ 2E_{\pm}$  & $2n\pi \pm cos^{-1}\left( \sqrt{\frac{2}{3}}\right) $ & $0$ & $1-4E_{\pm}^{2}$ & $\frac{1}{E_{\pm}}$ & $V_{1}, V_{3}$  \\  \hline
          $er4_{\pm}$ & $- 2E_{\pm}$  & $(2n+1)\pi \pm cos^{-1}\left(- \sqrt{\frac{2}{3}}\right) $ & $0$ & $1-4E_{\pm}^{2}$ & $\frac{1}{E_{\pm}}$ & $V_{1}, V_{3}$  \\  \hline
           $er5_{\pm}$ &$\pm \frac{2 \alpha _{2}}{\alpha _{1}}$ & $ \left(n\pi \pm cos^{-1}\left(\pm \sqrt{\frac{2}{3}}\right) \right)$ & $0$ & $1-\frac{4 \alpha _{2}^{2}}{\alpha _{1}^{2}}$ & $-\frac{\alpha _{1}}{\alpha _{2}}$  & $V_{2}$  \\  \hline
            $er6_{\pm}$ &$\pm \frac{2 \alpha _{1}}{\alpha _{0}}$ & $ \left(n\pi \pm cos^{-1}\left(\pm \sqrt{\frac{2}{3}}\right) \right)$ & $0$ & $1-\frac{4 \alpha _{1}^{2}}{\alpha _{0}^{2}}$ & $-\frac{\alpha _{0}}{\alpha _{1}}$  & $V_{4}$  \\  \hline
          $em1_{\pm}$ & $\sqrt{3}E_{\pm}$ & $ 2n\pi \pm \frac{\pi}{4}$ & $C_{\pm}$ & $0$ & $A_{\mp}$  & $V_{1}, V_{3}$  \\  \hline
          $em2_{\pm}$ & $-\sqrt{3}E_{\pm}$  & $ 2n\pi \pm \frac{3\pi}{4} $ & $C_{\pm}$ & $0$ & $A_{\mp}$ & $V_{1}, V_{3}$  \\  \hline
           $em3_{\pm}$ & $\pm \sqrt{3} E_{\pm}$ & $n\pi \pm \frac{\pi}{4} $ & $1-3E_{\pm}^{2}$ & $0$ & $\frac{1}{E_{\pm}}$  & $V_{1}, V_{3}$  \\  \hline
            $em4$ &$ \frac{\sqrt{3} \alpha _{2}}{\alpha _{1}}$ &  $2 n \pi \pm \frac{\pi}{4}$ & $1-\frac{3 \alpha _{2}^{2}}{\alpha _{1}^{2}}$ & $0$ & $-\frac{\alpha _{1}}{\alpha _{2}}$  & $V_{2}$  \\  \hline
            $em5$ &$ -\frac{\sqrt{3} \alpha _{2}}{\alpha _{1}}$ &  $2 n \pi \pm \frac{3\pi}{4}$ & $1-\frac{3 \alpha _{2}^{2}}{\alpha _{1}^{2}}$ & $0$ & $-\frac{\alpha _{1}}{\alpha _{2}}$  & $V_{2}$  \\  \hline
           $em6$ &$ \frac{\sqrt{3} \alpha _{1}}{\alpha _{0}}$ &  $2 n \pi \pm \frac{\pi}{4}$ & $1-\frac{3 \alpha _{1}^{2}}{\alpha _{0}^{2}}$ & $0$ & $-\frac{\alpha _{0}}{\alpha _{1}}$  & $V_{4}$  \\  \hline
            $em7$ &$ -\frac{\sqrt{3} \alpha _{1}}{\alpha _{0}}$ &  $2n \pi \pm \frac{3\pi}{4}$ & $1-\frac{3 \alpha _{1}^{2}}{\alpha _{0}^{2}}$ & $0$ & $-\frac{\alpha _{0}}{\alpha _{1}}$  & $V_{4}$  \\  \hline
          $ec1$ &  $r$ & $n\pi \pm\frac{\pi}{2}$ & $0$ & $0$ & $0$  & $V_{1}- V_{8}$ \\  \hline
          $ec2_{\pm}$ & $\pm \frac{1}{\sqrt{2}}E_{\pm}D_{\mp}$ &  \vtop{\hbox{\strut $n \pi ,$}\hbox{\strut $2n\pi \pm cos^{-1}\left( \frac{D_{\mp}}{2\sqrt{3}} \right)$}} & $0$ & $0$ & $A_{\mp}$  & $V_{1}, V_{3}$  \\  \hline
         \end{tabular}
         }
    \caption{ The list of equilibrium points for the system of equations given in Eq.$(\ref{polar_autonomous:1})$. The complete expressions of $A_\pm, B_\pm, C_\pm, D_\pm E_\pm$ are given in Appendix-\ref{app:eigenvalues}. The last column of the table shows the classes of potentials for which the fixed points exist.}
    \label{tab:fixedpoints}
\end{table*}

\begin{table*}[!hbt]
    \centering
    \scalebox{1.15}{
    \begin{tabular}{|c|c|}
       \hline {Equilibrium points} & { eigenvalues} \\ \hline
        $r$ &   \vtop{\hbox{\strut $4, 3, -1, 1, 0$ for $\theta=n\pi$;}\hbox{\strut   $4, -3, 2, 1, 0$ for $\theta=n\pi \pm \frac{\pi}{2}$} }  \\  \hline
        $m$ &   \vtop{\hbox{\strut  $3,3,-\frac{3}{2},-1,0$ for $\theta=n\pi$;}\hbox{\strut   $-3,3,\frac{3}{2},-1,0$ for $\theta=n\pi \pm \frac{\pi}{2}$} }  \\  \hline
        $e1$ &  \vtop{\hbox{\strut $6,3,3,2,-\sqrt{6} \alpha _{1}$, for even $n$;}\hbox{\strut $6,3,3,2,+\sqrt{6} \alpha _1$, for odd $n$} } \\  \hline
         $e2$ & $-4,-3,0,\frac{1}{2} \left(-3 + \sqrt{9-12 \alpha _{0}}\right),\frac{1}{2} \left(-3 - \sqrt{9-12 \alpha _{0}}\right)$ \\  \hline
          $e3$ &  \vtop{\hbox{\strut $2,3,6,\mp \sqrt{6}\sqrt{\alpha _{1}^2 - 4 \alpha _{0} \alpha _{2}} ,3-\sqrt{\frac{3}{2}}A_{\pm }$, for even $n$;}\hbox{\strut $2,3,6,\pm \sqrt{6}\sqrt{\alpha _{1}^2 - 4 \alpha _{0} \alpha _{2}} ,3+\sqrt{\frac{3}{2}}A_{\pm }$, for odd $n$} } \\  \hline
            $e4_{\pm}$ & $2,3,6,\mp \sqrt{6}\sqrt{\alpha _1^{2}-4 \alpha _0 \alpha _2} ,3-\sqrt{\frac{3}{2}}A_{\pm }$ \\  \hline
            $c$ &   \vtop{\hbox{\strut  $-4, -3, -3, 3, 0$ for $\theta=n\pi$;}\hbox{\strut    $-4,-3,-3,0,0$ for $\theta=n\pi \pm \frac{\pi}{2}$} }  \\  \hline
          $er1_{\pm}, er2_{\pm}$  & \vtop{\hbox{\strut $1,4,-8 \alpha _{2} -4 \alpha _{1}E_{\pm },$}\hbox{\strut $\frac{1}{2} \left(-1\pm \frac{1}{\alpha _{0}}\sqrt{- \left(15 \alpha _{0}^2 + 64 \alpha _{0} \alpha _{2} + 64 \alpha _{1} \alpha _{2} A_{\pm }\right)}\right),$  }}  \\  \hline
           $er3_{\pm}, er4_{\pm}$  & -  \\  \hline
           $er5_{\pm}$  &  $1,4,-4 \alpha _{2}, \frac{1}{2}\pm \left( \frac{1}{6\alpha_{1}}\right)\sqrt{3(3-8\sqrt{6})\alpha_{1}^{2}+96\sqrt{6} \alpha_{2}^{2}}$ \\  \hline
           $er6_{\pm}$  &  $1,4,\frac{4\alpha_{1}^{2}}{\alpha_{0}}, \frac{1}{2}\pm \left( \frac{1}{6\alpha_{0}}\right)\sqrt{3(3-8\sqrt{6})\alpha_{0}^{2}+96\sqrt{6} \alpha_{1}^{2}}$ \\  \hline
          $em1_{\pm},em2_{\pm}$ & \vtop{\hbox{\strut $-1,3,-6 \alpha _{2} -3 \alpha _{1}E_{\pm },$}\hbox{\strut $\frac{3}{4} \left(-1\pm \frac{1}{\alpha _{0}} \sqrt{- \left(7 \alpha _{0}^2 + 24 \alpha _{0} \alpha _{2} + 24 \alpha _{1} \alpha _{2} A_{\pm }\right)}\right),$ }}  \\  \hline
          $em3_{\pm}$ & - \\  \hline
          $em4,em5$ & $-1$, $3$, $-3\alpha_{2}$, $\frac{3}{4}\left( -1\pm \left( \frac{3}{4\alpha_{1}} \right)\sqrt{24\alpha_{2}^{2}-7\alpha_{1}^{2}} \right)$ \\  \hline
           $em6,em7$ & $-1$, $3$, $\frac{3\alpha_{1}^{2}}{\alpha_{0}}$, $\frac{3}{4}\left( -1\pm \left( \frac{3}{4\alpha_{0}} \right)\sqrt{24\alpha_{1}^{2}-7\alpha_{0}^{2}} \right)$ \\  \hline
          $ec1$ &  $-4,-3,0,\frac{1}{2} \left(-3\pm \sqrt{9-12 \alpha _{0} r_{c}^{2}}\right)$ \\  \hline
          $ec2_{\pm}$  & \vtop{\hbox{\strut $2,3,6,3\mp \frac{\sqrt{3}}{2} \sqrt{-\frac{2 \alpha _{0}}{\alpha _{2}}-\frac{2 \alpha _{1}}{\alpha _{2}} A_{\mp }},$}\hbox{\strut $\pm \frac{\sqrt{3} \alpha _{2}}{\alpha _{0}} A_{\pm }\sqrt{\alpha _{1}^2 - 4 \alpha _{0} \alpha _{2}} \sqrt{-\frac{2 \alpha _{0}}{\alpha _{2}}-\frac{2 \alpha _{1}}{\alpha _{2}} A_{\mp }}$, for even $n$ and $sin\theta =0$;}\hbox{\strut $2,3,6,3\pm\frac{\sqrt{3}}{2} \sqrt{-\frac{2 \alpha _{0}}{\alpha _{2}}-\frac{2 \alpha _{1}}{\alpha _{2}} A_{\mp }},$}\hbox{\strut $\mp \frac{\sqrt{3} \alpha _{2}}{\alpha _{0}} A_{\pm }\sqrt{\alpha _{1}^2 - 4 \alpha _{0} \alpha _{2}} \sqrt{-\frac{2 \alpha _{0}}{\alpha _{2}}-\frac{2 \alpha _{1}}{\alpha _{2}} A_{\mp }}$, for odd $n$ and $sin\theta =0$;}\hbox{\strut $2\alpha _{0} +\alpha _{1} A_{\mp },-\frac{\alpha _{0}}{\alpha _{2}} -\left(\frac{\alpha _{1}}{\alpha _{2}}\right) A_{\mp },-\frac{1}{\alpha _{2}} \left( \alpha _{0} + 3 \alpha _{2} \right) -\left(\frac{\alpha _{1}}{\alpha _{2}}\right) A_{\mp },$}\hbox{\strut $-\frac{1}{\alpha _{2}} \left( \alpha _{0}+4 \alpha _{2} \right) -\left(\frac{\alpha _{1}}{\alpha _{2}}\right) A_{\mp },-\frac{1}{2 \alpha _{2}} \left( \alpha _{0}+6 \alpha _{2} \right) -\left(\frac{\alpha _{1}}{2 \alpha _{2}}\right) A_{\mp }$, for $cos\theta =\frac{D_{\mp}}{2\sqrt{3}}$}}  \\  \hline
         \end{tabular}
         }
    \caption{ The list of eigenvalues of the equilibrium points given in Table-$\ref{tab:fixedpoints}$. For the fixed points $er3_\pm, er4_\pm, em3_\pm$ the analytical forms of the eigenvalues are complicated and hence not included in the table. These eigenvalues can be computed for some particular choices of $\alpha$ parameters, see the text for more details. }
    \label{tab:eigenvalues}
\end{table*}

\begin{table*}[!hbt]
    \centering
    \scalebox{0.95}{
    \begin{tabular}{|c|c|c|c|}
       \hline {Equilibrium points} & { $q$} & {$\omega_{\phi}$} & {$\omega_{tot}$} \\ \hline
        $r$ &  $1$ & \vtop{\hbox{\strut $1$ for $\theta=n\pi$;}\hbox{\strut $-1$ for $\theta=n\pi \pm \frac{\pi}{2}$} }& $\frac{1}{3}$ \\  \hline
        $m$ &  $\frac{1}{2}$  &  \vtop{\hbox{\strut $1$ for $\theta=n\pi$;}\hbox{\strut $-1$ for $\theta=n\pi \pm \frac{\pi}{2}$} } & $0$ \\  \hline
        $e1$,$e3$, $e4_{\pm}$ &  $2$ &  $1$  & $1$ \\  \hline
           $e2$ &  $-1$ &  $-1$  & $-1$ \\  \hline
            $c$ &  $-1$ &  \vtop{\hbox{\strut $1$ for $\theta=n\pi$;}\hbox{\strut $-1$ for $\theta=n\pi \pm \frac{\pi}{2}$} }  & $-1$ \\  \hline
         \vtop{\hbox{\strut $er1_{\pm}$, $er2_{\pm}$,}\hbox{\strut $er3_{\pm}$, $er4_{\pm}$}}  & \vtop{\hbox{$\frac{\alpha _0^2+16 \alpha _2 \alpha _0+8 \left(\sqrt{\alpha _1^2-4 \alpha _0 \alpha _2}-\alpha _1\right) \alpha _1}{\alpha _0^2}$,}\hbox{ for $er_{+}$}\hbox{ $\frac{\alpha _0^2+16 \alpha _2 \alpha _0-8 \alpha _1 \left(\sqrt{\alpha _1^2-4 \alpha _0 \alpha _2}+\alpha _1\right)}{\alpha _0^2}$,}\hbox{ for $er_{-}$ }}  &   $\frac{1}{3}$  & $\frac{1}{3}$ \\  \hline
            $er5_{\pm}$,  $er6_{\pm}$  & \vtop{\hbox{\strut $1-\frac{16 \alpha _1^2}{\alpha _0^2}$, for $er5_{\pm}$;}\hbox{\strut $1-\frac{16 \alpha _2^2}{\alpha _1^2}$, for $er6_{\pm}$} }  &   $\frac{1}{3}$  & $\frac{1}{3}$ \\  \hline
          \vtop{\hbox{\strut $em1_{\pm}$,} \hbox{\strut$em2_{\pm}$, $em3_{\pm}$ }} & \vtop{\hbox{$\frac{\alpha _0^2+18 \alpha _2 \alpha _0+9 \left(\sqrt{\alpha _1^2-4 \alpha _0 \alpha _2}-\alpha _1\right) \alpha _1}{2 \alpha _0^2}$,}\hbox{ for $em_{+}$}\hbox{ $\frac{\alpha _0^2+18 \alpha _2 \alpha _0-9 \alpha _1 \left(\sqrt{\alpha _1^2-4 \alpha _0 \alpha _2}+\alpha _1\right)}{2 \alpha _0^2}$,}\hbox{ for $em_{-}$ }}  &   $\frac{1}{3}$  & $\frac{1}{3}$ \\  \hline
           $em4, em5$ &$\frac{1}{2}-\frac{9 \alpha _2^2}{\alpha _1^2}$ &   $0$ & $ 0$ \\  \hline
            $em6, em7$ & $\frac{1}{2}-\frac{9 \alpha _1^2}{\alpha _0^2}$ &   $0$ & $ 0$ \\  \hline
            $ec1$ & $-1$ &  $-1$  & $-1$ \\  \hline
           $ec2_{\pm}$  &  \vtop{\hbox{\strut $-4$, for $\theta =n\pi$;}\hbox{\strut $\frac{2 \left(\alpha _0-2 \alpha _2\right) \alpha _2-\alpha _1 \left(\sqrt{\alpha _1^2-4 \alpha _0 \alpha _2}+\alpha _1\right)}{4 \alpha _2^2}$,}\hbox{ for $r_{c}=\pm \frac{1}{\sqrt{2}}E_{+}D_{-}$}\hbox{\strut $\frac{\left(\sqrt{\alpha _1^2-4 \alpha _0 \alpha _2}-\alpha _1\right) \alpha _1+2 \alpha _2 \left(\alpha _0-2 \alpha _2\right)}{4 \alpha _2^2}$,}\hbox{ for $r_{c}=\pm \frac{1}{\sqrt{2}}E_{-}D_{+}$} }  &  \vtop{\hbox{\strut $1$, for $\theta =n\pi$;}\hbox{\strut $\frac{\alpha _1 \left(\sqrt{\alpha _1^2-4 \alpha _0 \alpha _2}+\alpha _1\right)-2 \alpha _2 \left(\alpha _0+3 \alpha _2\right)}{6 \alpha _2^2}$,}\hbox{ for $r_{c}=\pm \frac{1}{\sqrt{2}}E_{+}D_{-}$}\hbox{\strut $\frac{\alpha _1^2-\sqrt{\alpha _1^2-4 \alpha _0 \alpha _2} \alpha _1-2 \alpha _2 \left(\alpha _0+3 \alpha _2\right)}{6 \alpha _2^2}$,}\hbox{ for $r_{c}=\pm \frac{1}{\sqrt{2}}E_{-}D_{+}$} }  &  \vtop{\hbox{\strut $1$, for $\theta =n\pi$;}\hbox{\strut $\frac{\alpha _1 \left(\sqrt{\alpha _1^2-4 \alpha _0 \alpha _2}+\alpha _1\right)-2 \alpha _2 \left(\alpha _0+3 \alpha _2\right)}{6 \alpha _2^2}$,}\hbox{ for $r_{c}=\pm \frac{1}{\sqrt{2}}E_{+}D_{-}$}\hbox{\strut $\frac{\alpha _1^2-\sqrt{\alpha _1^2-4 \alpha _0 \alpha _2} \alpha _1-2 \alpha _2 \left(\alpha _0+3 \alpha _2\right)}{6 \alpha _2^2}$,}\hbox{ for $r_{c}=\pm \frac{1}{\sqrt{2}}E_{-}D_{+}$} }   \\  \hline
         \end{tabular}
         }
    \caption{ The values of cosmological parameters for the equilibrium points given in Table-$\ref{tab:fixedpoints}$.}
    \label{tab:cosmoparameters}
\end{table*}

\section{Stability Analysis}
The equilibrium points of the system of equations given in Eq.$(\ref{polar_autonomous:1})$ are listed in Table-$\ref{tab:fixedpoints}$, while the eigenvalues corresponding to these points are presented in Table-$\ref{tab:eigenvalues}$. The equilibrium points are calculated by solving the simultaneous system of equations: $r^{'}=0$, $\theta^{'}=0$, $\Omega_{m}^{'}=0$, $\Omega_{r}^{'}=0$, and $\lambda^{'}=0$. The existence of equilibrium points for the corresponding potentials is also given in the last column of Table-$\ref{tab:fixedpoints}$. The values for the cosmological parameters, such as the deceleration parameter, the equation of state (EoS) of the scalar field, and the total EoS for each fixed point, are provided in Table-\ref{tab:cosmoparameters}.

In this context, the subscript `$c$' is used to denote the equilibrium points $(r_{c},\hspace{0.1cm} \theta_{c}, \hspace{0.1cm}\Omega_{mc},\hspace{0.1cm} \Omega_{rc},\hspace{0.1cm} \lambda_{c})$. Subsequently, we analyze the stability of these equilibrium points by categorizing them according to the different epochs of the universe they represent.

\subsection{Radiation Dominated Era} 

The equilibrium points indicating the radiation domination era are characterised by $\Omega_{rc}=1$ and represented by `$r$' which are of the nonisolated type. For all the equilibrium points of this era, the corresponding Jacobian matrix has two different sets of eigenvalues depending on whether $\theta_{c}=n\pi$ or $\theta_{c}=n\pi \pm \frac{\pi}{2}$. In both cases, the Jacobian matrix has both positive and negative eigenvalues. So, these equilibrium points are saddle-like in nature and exist for all potentials ($V_{1}-V_{8}$) listed in Table-$\ref{tab:potentials}$. From Table-\ref{tab:cosmoparameters} it can be seen as expected that at these equilibrium points the universe is decelerating, and $w_{tot} = 1/3$.

\subsection{Matter Dominated Era}

The equilibrium points indicating the matter dominated era are defined by $\Omega_{mc}=1$, and are denoted by `$m$'. The Jacobian matrix, corresponding to any equilibrium point during this era, exhibits two distinct sets of eigenvalues depending on whether $\theta_{c}=n\pi$ or $\theta_{c}=n\pi \pm \frac{\pi}{2}$. For each $n\in \mathbb{Z}$, the equilibrium points during this period are categorized as saddle points due to the presence of both positive and negative eigenvalues corresponding to these fixed points. These equilibrium points for the matter domination era are present for all potentials $V_{1} - V_{8}$ and represent a decelerating universe with $w_{tot}=0$.

\subsection{Scalar field dominated era}

For the equilibrium points in the scalar field domination era, we have $r_{c}= 1$. In this case, the equilibrium points are given by `$e1$' - `$e4_{\pm}$'. The eigenvalues of the Jacobian matrix at `$e1$' depend on $\alpha_{1}$. But, for the existence of positive eigenvalues of the Jacobian matrix at `$e1$', the equilibrium points given by `$e1$' are unstable and exist for the potentials $V_{2}$ and $V_{6} - V_{8}$. These equilibrium points represent a decelerating universe with $w_{tot}=1$.

\begin{figure}[ht]
            \centering
            \includegraphics[width=\columnwidth]{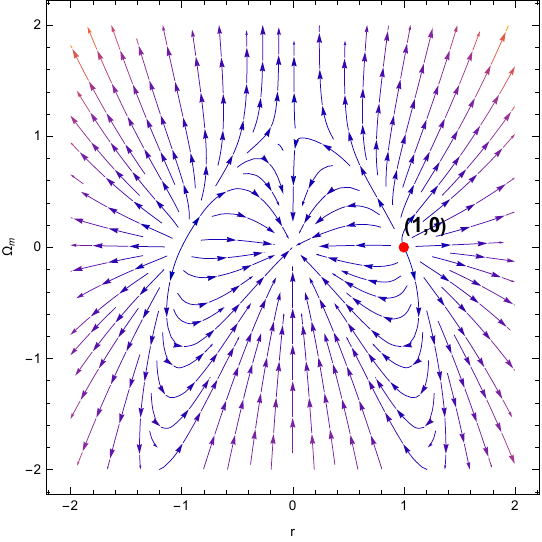}
            \caption{Phase plot for the system of equations Eq.$(\ref{polar_autonomous:1})$ in ($r -\Omega_{m}$) plane for the fixed point `$e_2$'.}
            \label{fig:pplote2}
    \end{figure}

The equilibrium points indicated by `$e2$' are nonhyperbolic, characterised by a zero eigenvalue. These fixed points may be stable within the parameter range $0<\alpha_0 \leq \frac{3}{4}$, as the remaining two eigenvalues can be negative. Due to their non-hyperbolic nature, linear stability analysis cannot determine their stability. Instead, stability must be assessed using center manifold theory or by numerically plotting the system's phase around the fixed point. Given that the system's dimension exceeds three, a complete phase plot is impractical; hence, phase portraits are drawn on various projected planes. If all projected phase planes depict the equilibrium point as stable, it is stable in the full space. Conversely, if any projected phase space shows instability, the equilibrium point is unstable in the entire space. For the equilibrium point `$e2$', we draw a phase plot (Fig.$\ref{fig:pplote2}$) in the ($r ,\Omega_{m}$) plane. This phase plot indicates that the equilibrium points specified by `$e2$' are unstable on the ($r ,\Omega_{m}$) plane and consequently in the full space. Those equilibrium points represent an accelerating universe with $w_{\phi}=w_{tot}=-1$. 

The equilibrium points, labelled as `$e3$' and `$e4_{\pm}$', are identified for $r_{c} = 1$, $\Omega_{mc} = 0$, $\Omega_{rc} = 0$, and $\lambda_{c} = A_{\pm}$ (refer to Appendix: \ref{app:eigenvalues} for the complete form of $A_{\pm}$). These fixed points exist for the potentials $V_{1} - V_{3}$ and $V_{7}$. However, the values of $\theta_{c}$ are not the same for these points. These equilibrium points exist when $\alpha_{2} \neq 0$ and $\alpha_{1}^{2} - 4\alpha_{0}\alpha_{2} \geq 0$. The eigenvalues of `$e3$' are determined by whether $n$ is even or odd, while the eigenvalues of `$e4_{\pm}$' are independent of $n$. In all cases, the Jacobian matrix evaluated at `$e3$' and `$e4_{\pm}$' exhibits positive eigenvalues. Consequently, the fixed points represented by `$e3$' and `$e4_{\pm}$' are unstable regardless of the values of $n$ and the parameters $\alpha_{0}, \alpha_{1}$, and $\alpha_{2}$. These fixed points represent a decelerating universe with $w_{\phi}=w_{tot}=1$.

\subsection{$\Lambda $ - Dominated Era}
\begin{figure*}[ht]
            \centering
            \includegraphics[width=\textwidth, height=1.2\textwidth]{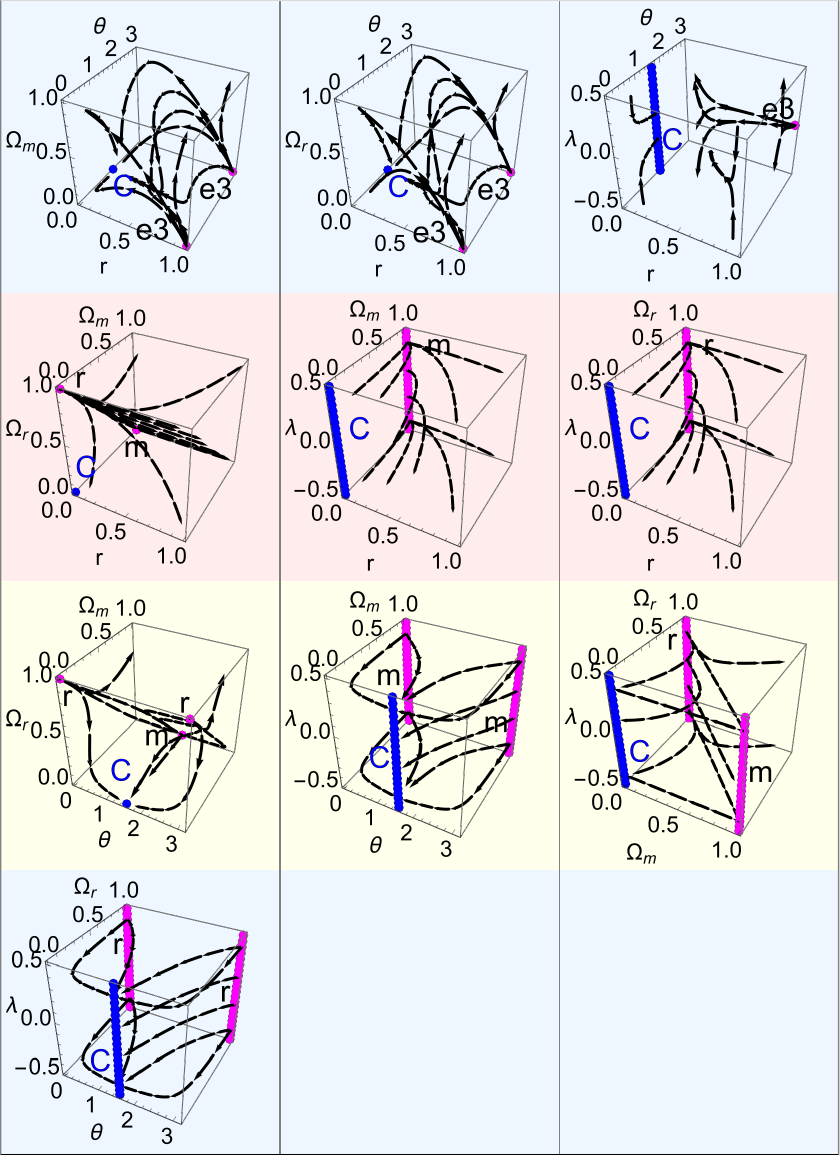}
            \caption{Phase plots in different $3D$ projected spaces for the equilibrium point `$c$' and for the potential $V_{2}$. The location of the fixed point `$c$' is depicted by the blue point/line. The points/line in magenta color represents the location of the other fixed points which are unstable in nature. }
            \label{fig:pplotcv2}
    \end{figure*}

The equilibrium points representing the cosmological constant or $\Lambda$-domination era are represented by $r_{c}=0$, $\Omega_{mc}=0$, and $\Omega_{rc}=0$ and exist for all classes of potentials. In this case, the equilibrium points are given by `$c$' in the Table-$\ref{tab:fixedpoints}$ and they are nonisolated equilibrium points. The eigenvalues of the Jacobian matrix evaluated at `$c$' depend on $\theta_{c}$. For $\theta_{c}=n\pi$ the Jacobian matrix has a mixture of positive and negative eigenvalues, hence the fixed points in this case are saddle.  On the other hand, finding the stability of the equilibrium points `$c$' for $\theta_{c}=n\pi \pm \frac{\pi}{2}$ is more involved since it has two zero eigenvalues and other all negative eigenvalues. Three negative eigenvalues correspond to a 3D stable manifold, whereas the two zero eigenvalues correspond to a 2D center manifold. In general, one can use the center manifold theorem to investigate the stability of this fixed point, rather here we draw the phase plots on 3D planes. All the points in red color and the points on the lines shown in red in the projected phase spaces, as depicted in Fig. \ref{fig:pplotcv2}, represent the equilibrium point `$c$' for the $V_2$ potential.  From this figure, we can see that the fixed points given by `$c$' are stable for $\theta = n\pi \pm \frac{\pi}{2}$ in all projected subspaces and hence stable in the entire 5D space. We have also checked that for the other class of potentials, the behaviour of the fixed point remains the same. These fixed points represent an accelerating universe with $w_{\phi}=1$ and $w_{tot}=-1$ for $\theta =n\pi$ whereas $w_{\phi}=w_{tot}=-1$ for $\theta =n\pi \pm \frac{\pi}{2}$. Interestingly, in the first case, the scalar field behaves as a stiff fluid. 

    \begin{figure}[ht]
            \centering
            \includegraphics[width=\columnwidth]{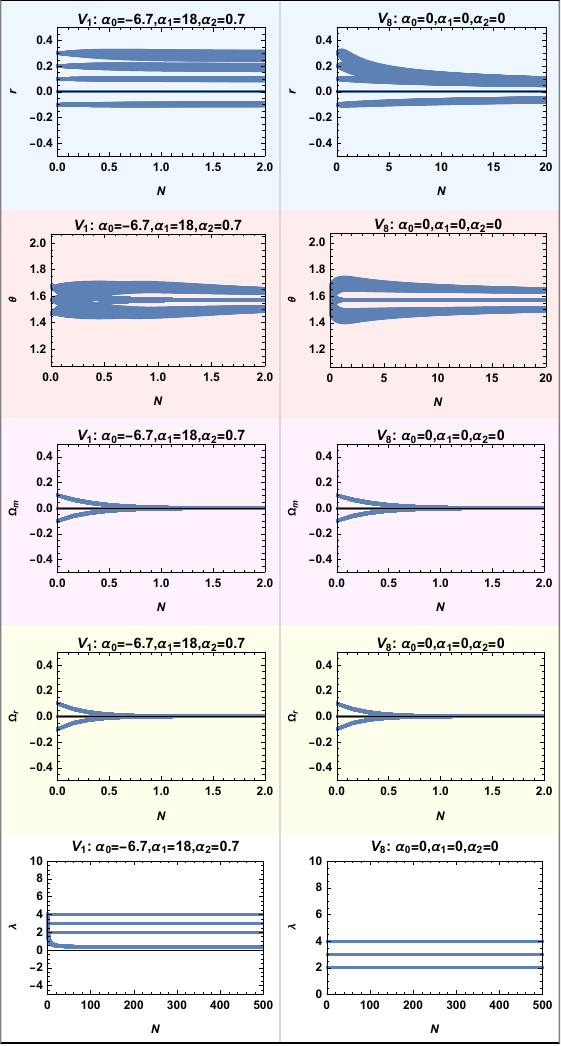}
            \caption{Plots of the numerical solutions for each dynamical variable by considering perturbations around the equilibrium point `$ec1$'. The plots in the left column are shown for potential $V_1$ and on the right it is shown for the potential $V_8$.}
            \label{fig:pltec1v1v2}
    \end{figure}

\subsection{The Era Shared by Scalar Field and other components}

In our setup, there are some specific fixed points that indicate scenarios where the energy budget is shared between the scalar field and other components. 

\subsubsection*{\textbf{The Era Shared by the Scalar Field and Radiation}}
The fixed points `$er1_{\pm}$' to `$er6_{\pm}$' represent the state where the universe's energy budget is shared by radiation and the scalar field by the relation $r_{c}^{2}+\Omega_{rc}=1$. For each $n \in \mathbb{Z}$, the equilibrium points `$er1_{\pm}$', `$er2_{\pm}$', `$er5_{\pm}$', and `$er6_{\pm}$' exhibit positive eigenvalues, making these points unstable or saddle depending on the choice of parameters $\alpha _{0}, \alpha _{1}, \alpha _{2}$. The eigenvalues of the Jacobian matrix at `$er3_{\pm}$' and `$er4_{\pm}$' are complicated in general, making it difficult to draw any conclusions about their stability. For these points, we evaluate the stability at specific values of $\alpha$ parameters. We have considered the same value of the $\alpha$ parameters that have been used to plot Fig.$\ref{fig:v1}$ and Fig.$\ref{fig:v3}$, for potentials $V_{1}$ and $V_{3}$ respectively. For the potential $V_{1}$ the eigenvalues at `$er3_{\pm}$' are $1, 4, -2.8394$, and $-0.5 \pm 1.93041 i$, while at `$er4_{\pm}$' they are $-201.793, -11.2262, 10.2262, 4$, and $1$, indicating that these points are saddle points. For potential $V_{3}$, the eigenvalues at `$er3_{\pm}$' and `$er4_{\pm}$' are $1, 4, -4.48$, and $-0.5 \pm 1.92544 i$, showing that they are saddle points. The fixed points `$er1_{\pm}$' and `$er2_{\pm}$' exist for the potentials $V_{1}$ and $V_{3}$; `$er5_{\pm}$' exist for the potential $V_{2}$; and `$er6_{\pm}$' for potential $V_{4}$. For all possible equilibrium points in this era $w_{\phi}=w_{tot}=1/3$ representing a decelerated universe.

\subsubsection*{\textbf{The Era Shared by the Scalar Field and Matter}}

For the equilibrium points in the era shared by scalar field and matter, we have $r_{c}^{2}+\Omega_{mc}=1$. In this case, the equilibrium points are given by `$em1_{\pm}$' - `$em7$'. For each $n\in \mathbb{Z}$, though the eigenvalues of the Jacobian matrix evaluated at the equilibrium points `$em1_{\pm}$', `$em2_{\pm}$' and `$em4$' - `$em7$' depend on the parameters $\alpha_{0}$, $\alpha_{1}$ and $\alpha_{2}$; but it has a positive eigenvalue irrespective of $\alpha_{i}$. So, the equilibrium points given in this era are unstable in nature. The equilibrium points `$em1_{\pm}$' - `$em3_{\pm}$' exist for the potentials $V_{1}$, $V_{3}$; `$em4$' and `$em5$' exist for the potential $V_{2}$; `$em6$' and `$em7$' exist for the potential $V_{4}$. The eigenvalues of the Jacobian matrix evaluated at the equilibrium point `$em3_{\pm}$' are very complicated. So, for `$em3_{\pm}$', we evaluate the stability similarly as the case for `$er3_{\pm}$' and `$er4_{\pm}$'. For the potential $V_{1}$ the eigenvalues at `$em3_{+}$' are $3, -1, -2.12955$, and $-0.75 \pm 1.97931 i$, while at `$er3_{-}$' they are $ 9.06334, 3, -1, -10.5633$, and $-151.345$, indicating that these points are saddle points. For potential $V_{3}$, the eigenvalues at `$em3_{\pm}$' are $3, -1, -3.36$, and $-0.75 \pm 1.97522 i$, showing that they are saddle points. The exact value of the deceleration parameter depends on $\alpha$ parameters for all the potentials belonging to this era. However, for all possible equilibrium points in this era $w_{\phi}=w_{tot}=0$ or $1/3$ representing a decelerated universe.

\subsubsection*{\textbf{The Era Shared by the Scalar Field and $\Lambda$}}

For the equilibrium points in the era shared by scalar field and $\Lambda$, we have $r_{c}^{2}+\Omega_{\Lambda_c }=\Omega_{\phi_c} + \Omega_{\Lambda_c}=1$. In this case, the equilibrium points are given by `$ec1$' and `$ec2_{\pm}$'. The equilibrium points given by `$ec1$' exist for all potentials $V_{1} - V_{8}$. The nature of the equilibrium points given by `$ec1$' depends on the parameter $\alpha_{0}$. If $0<\alpha_{0}<\frac{3}{4r_{c}^{2}}$, then the equilibrium points given by `$ec1$' are normally hyperbolic equilibrium points and the Jacobian matrix has four eigenvalues with negative real parts and one zero eigenvalue. Again, if $\alpha_{0}=0$ then the equilibrium points given by `$ec1$' are of non-hyperbolic type and the Jacobian matrix has three negative and two zero eigenvalues. To check the stability of those equilibrium points we plot the evolution of the dynamical variables $r, \theta, \Omega_{m}, \Omega_{r}$ and $\lambda$ by solving the system numerically where the system has been perturbed from the equilibrium point `$ec1$'. This approach for finding the stability of the non-hyperbolic fixed points has been previously used in \cite{Roy:2014yta,Roy:2014hsa,Roy:2017mnz}. In Fig.$\ref{fig:pltec1v1v2}$ for example we have plotted the evolution of the system with $N$ for the potentials $V_{1}$ (on the left column) and $V_{8}$ (on the right column) subjected to perturbations around the fixed point `$ec_1$'. We can see that the evolution of the dynamical variables does not diverge but rather stays in the neighbourhood of the fixed point, indicating its stable nature.  We have also checked that the qualitative behaviour of the fixed point remains the same even for the other potentials. Thus, the fixed points given by `$ec1$' are stable for $\alpha_{0}=0$. Again, if $\alpha_{0}<0$, then the equilibrium points given by `$ec1$' are non-hyperbolic type and the Jacobian matrix has a positive eigenvalue. Thus, the equilibrium points given by `$ec1$' are stable for $0\leq \alpha_{0}\leq \frac{3}{4r_{c}^{2}}$, where $0\leq r_{c}\leq 1$ and unstable for $\alpha_{0}<0$.  These fixed points represent an accelerating universe with $w_{\phi}=w_{tot}=-1$.

The equilibrium points `$ec2_{\pm}$' are defined by $r_{c}=\pm \frac{1}{\sqrt{2}}E_{\pm}D_{\mp}$, $\Omega_{mc}=0$, $\Omega_{rc}=0$, and $\lambda_{c}=A_{\mp}$ and exists for potentials $V_1$ and $V_3$. There are two distinct types of $\theta_{c}$ at these points: $\theta_{c}=n\pi$ and $\theta_{c}=2n\pi \pm \cos^{-1}\frac{D_{\mp}}{2\sqrt{3}}$. For $\theta_{c}=n\pi$, the eigenvalues of the Jacobian matrix vary depending on whether $n$ is even or odd, but some eigenvalues are always positive, hence the fixed points are unstable. For $\theta_{c}=2n\pi \pm \cos^{-1}\frac{D_{\mp}}{2\sqrt{3}}$, the form of the eigenvalues is complicated, and it is not possible to conclude stability analytically for this fixed point for any general choice of the parameters $\alpha$. Thus, stability was assessed for specific values of the parameters $\alpha_{0}$, $\alpha_{1}$, and $\alpha_{2}$ that are considered for the numerical solutions shown in Fig.$\ref{rsh01}$ and Fig.$\ref{rsh03}$. The Jacobian matrix has at least one positive eigenvalue for our choices of the $\alpha$ parameters, making `$ec2_{\pm}$' unstable. These points also correspond to an accelerating universe.

\section{Numerical investigation}
In our analysis of the phase space for the current model, we observed that there is no late-time attractor completely dominated by the scalar field. The only possible late-time attractors are either completely dominated by the cosmological constant or a scenario where there is only the cosmological constant and scalar field present in the universe; the rest of the components are absent. Although there might be some epochs where the scalar field and other components coexist, represented by saddle or unstable fixed points.  For example, the fixed points in Table:\ref{tab:fixedpoints} where $r_c$ and some other components like $\Omega_{mc}$ or $\Omega_{rc}$ are non-zero. In the following, we have numerically evolved the system to investigate the evolution of the different cosmological parameters, and we will see, as expected from the fixed point analysis, that there are possibilities of some eras of the universe which are shared by the scalar field and other components of the universe much before the standard era of dark energy domination.

We have numerically solved the system of equations Eq.$(\ref{polar_autonomous:1})$ by considering the specific initial conditions given in Table-$\ref{tab:parameterspace}$. In particular, except for the initial condition of $\lambda$ for all potentials, the initial conditions for the other variables remain almost the same. The initial condition is set before the radiation-matter equality at $N=-10.1$, corresponding to $z\approx 22026$, deep in the radiation-dominated era. The range of $\alpha$ parameters and $\lambda_{ini}$ for each class of potential is presented in Table-$\ref{tab:potentials}$, for which the qualitative behaviour of our analysis remains the same.

\begin{figure}[h!]
            \centering
        \includegraphics[width=\columnwidth]{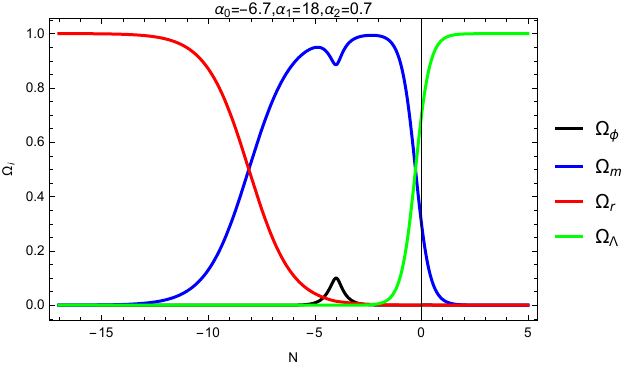}
            \caption{Plot of the density parameters for different components of the universe for the potential $V_{1}$. }
            \label{fig:v1}
    \end{figure}

\begin{figure}[ht]
            \centering
            \includegraphics[width=\columnwidth]{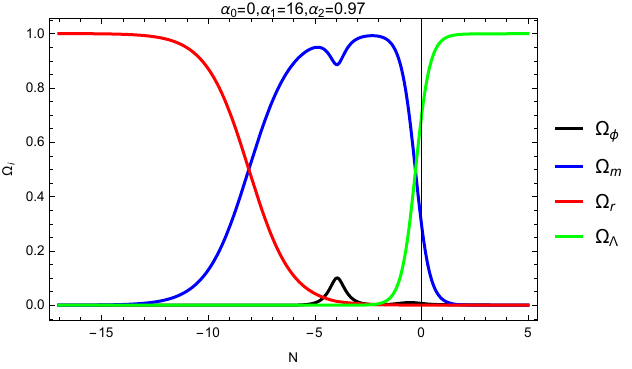}
            \caption{Plot of the density parameters for different components of universe for the potential $V_{2}$.}
            \label{fig:v2}
    \end{figure}

\begin{figure}[ht]
            \centering
            \includegraphics[width=\columnwidth]{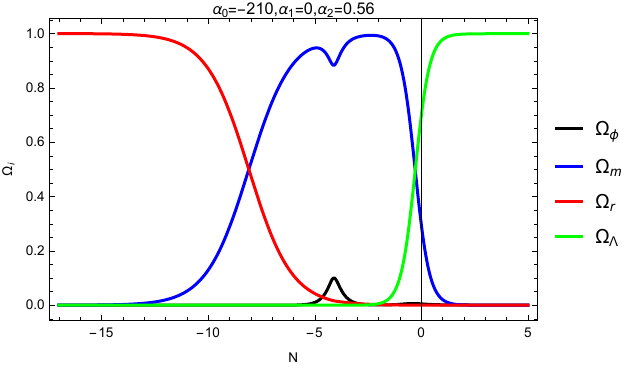}
            \caption{Plot of the density parameters for different components of universe for the potential $V_{3}$. }
            \label{fig:v3}
    \end{figure}

\begin{figure}[ht]
            \centering
            \includegraphics[width=\columnwidth]{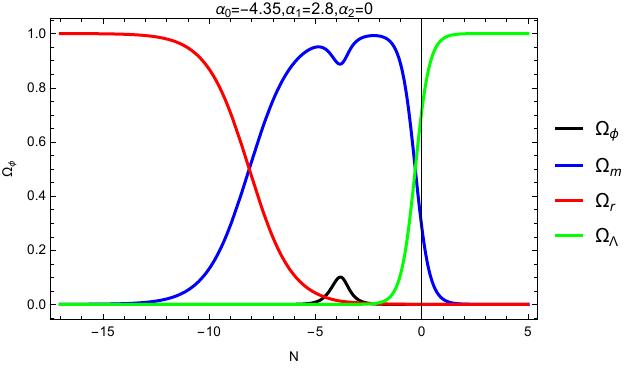}
            \caption{Plot of the density parameters for different components of universe for the potential $V_{4}$.}
            \label{fig:v4}
    \end{figure}

\begin{figure}[ht]
            \centering
            \includegraphics[width=\columnwidth]{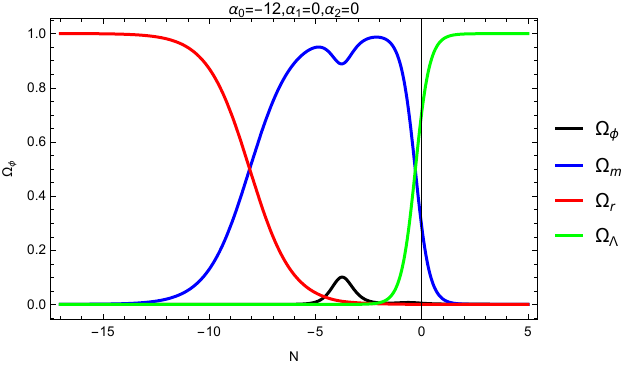}
            \caption{Plot of the density parameters for different components of universe for the potential $V_{5}$. }
            \label{fig:v5}
    \end{figure}

\begin{figure}[ht]
            \centering
            \includegraphics[width=\columnwidth]{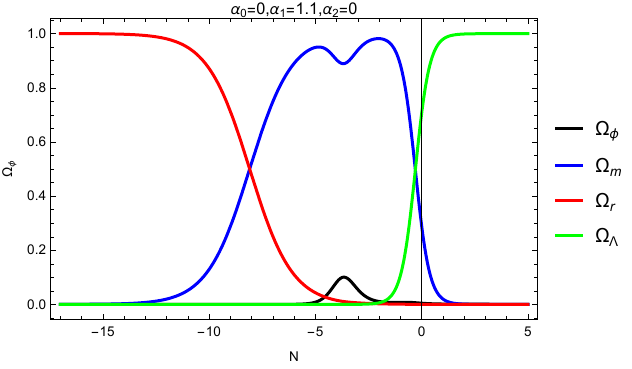}
            \caption{Plot of the density parameters for different components of universe for the potential $V_{6}$.}
            \label{fig:v6}
    \end{figure}

\begin{figure}[ht]
            \centering
            \includegraphics[width=\columnwidth]{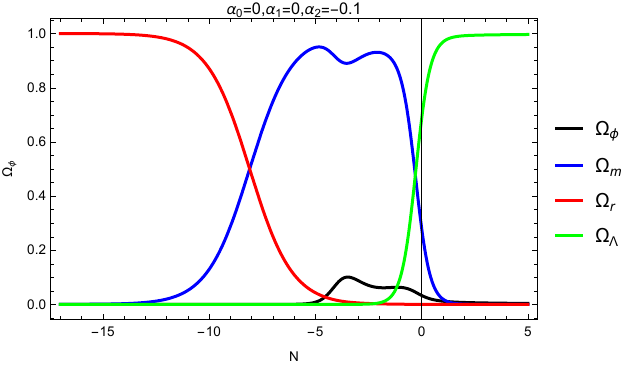}
            \caption{Plot of the density parameters for different components of universe for the potential $V_{7}$. }
            \label{fig:v7}
    \end{figure}

\begin{figure}[ht]
            \centering
            \includegraphics[width=\columnwidth]{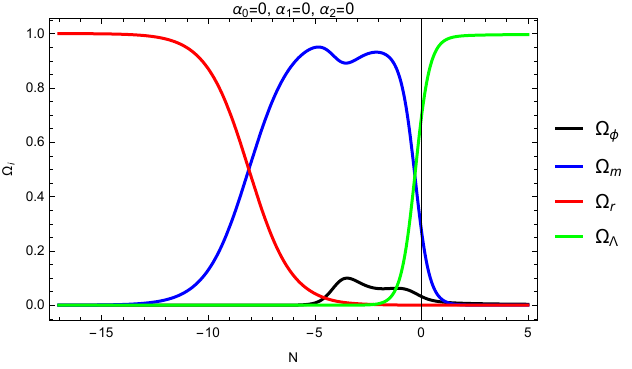}
            \caption{Plot of the density parameters for different components of universe for the potential $V_{8}$.}
            \label{fig:v8}
    \end{figure}

The evolution of the density parameters for various components of the universe corresponding to all the classes of potentials are shown in Fig.\ref{fig:v1} to  Fig.\ref{fig:v8} respectively. For the classes of potentials, $V_{1}$ - $V_{6}$ one can see the scalar field contribute to the energy budget of the universe earlier than the expected dark energy domination epoch for a short period resembling EDE-like behaviour of the scalar field appearing naturally in the deep-matter dominated era, which is almost $10\%$ of the total energy budget of the universe. But, at present the contribution of the scalar field is negligible for these potentials. Whereas for the classes of potentials, $V_{7}$ and $V_{8}$ also the similar behaviour of the scalar field appears naturally in the deep matter-dominated era, which is almost $10\%$ of the total energy budget of the universe. Contrary to the previous case, for these two classes of potentials, at present as well as near future, the scalar field has a non-negligible contribution to the total energy budget of the universe. The redshift parameter values for matter-radiation equality and the occurrence of this behavior of the scalar field are provided in Table-$\ref{tab:parameterspace}$  for all classes of potentials.

\begin{figure}[ht]
            \centering
            \includegraphics[width=\columnwidth]{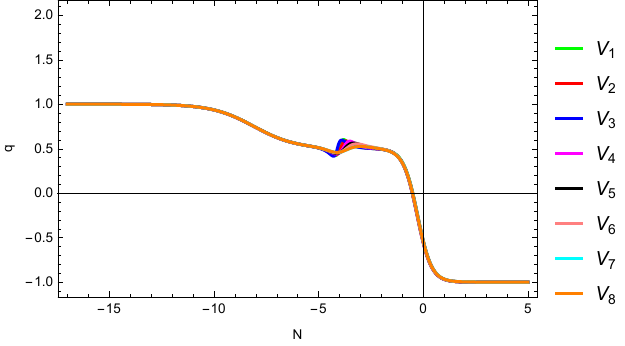}
            \caption{Plots of the deceleration parameter with respect to different  classes of potentials.}
            \label{fig:deceleration}
    \end{figure}

\begin{figure}[ht]
            \centering
            \includegraphics[width=\columnwidth]{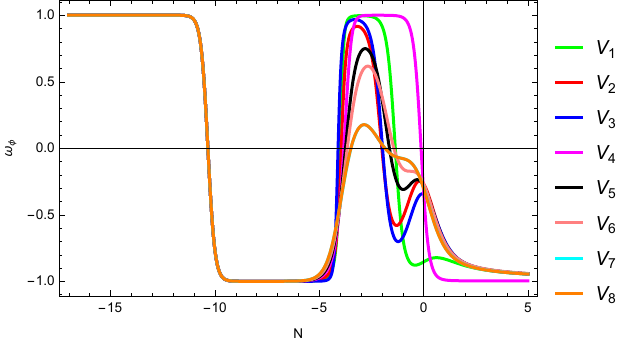}
            \caption{Plots of the equation of state for scalar field for different classes of potentials.}
            \label{fig:omegaphi}
    \end{figure}

\begin{figure}[ht]
            \centering
            \includegraphics[width=\columnwidth]{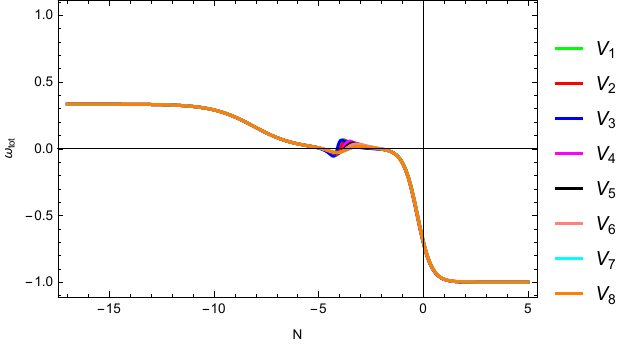}
            \caption{Plots of the total equation of state with respect to different classes of potentials.}
            \label{fig:omegatotal}
    \end{figure}

We have also plotted the evolution of different cosmological parameters like the deceleration parameter in Fig.$\ref{fig:deceleration}$, the equation of state of the scalar field in Fig.$\ref{fig:omegaphi}$ and the total equation of state in Fig.$\ref{fig:omegatotal}$ for all potentials. Those plots are drawn for the same values of the parameters $\alpha$ for which Fig.\ref{fig:v1} to  Fig.\ref{fig:v8} are drawn. For the evolution of the deceleration parameter, one can see that its value has become negative in the recent past from positive, showing the accelerating expansion of the universe. It can be seen that the transition of the universe from the decelerated to the accelerated phase remains smooth throughout except for the epoch where the scalar field shows early time contribution. During this epoch, the deceleration parameter shows some increments suggesting more deceleration of the universe due to the activation of the scalar field. Later, it started to decrease and become negative with time. From the plot of the EOS of the scalar field in Fig.$\ref{fig:omegaphi}$ one can see that in the very early universe, the scalar field behaves like a stiff fluid with EOS $w_{\phi}=1$ then it becomes $w_{\phi}=-1$. During the early contribution period of the scalar field, it increases and becomes positive and the scalar field does not contribute to the acceleration of the universe since $w_\phi > -1/3$ rather it was slowing down the expansion of the universe. This can also be seen from the plot of the $w_{tot}$ in Fig.$\ref{fig:omegatotal}$ since during this epoch the $w_{tot}$ becomes positive and starts to decrease again. 

\begin{figure}[ht]
            \centering
            \includegraphics[width=\columnwidth]{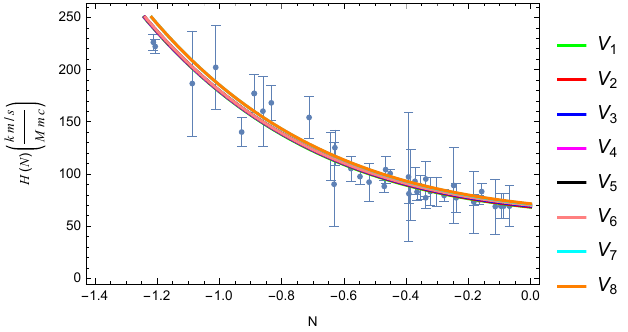}
            \caption{Plot of the evaluation of the Hubble parameter $H(N)$ versus $N$ for different classes of potentials. The data from various observations are also included for comparison. }
            \label{fig:Hparam}
    \end{figure}

In Fig.\ref{fig:Hparam} we have shown the evolution of the $H(z)$ vs $z$ for all the classes of potentials together with the observed data from different observations for comparison. Please see Section:\ref{sec:data} for more details about the data used in this plot.  It can be seen that even though the scalar field contributed during the matter-dominated era, these potentials fit the data quite well, at least during the late time evolution.

A similar approach to study the early dark energy behavior of the scalar field using dynamical system analysis was used in \cite{Copeland_2024, Ramadan:2023ivw}. Both works show that EDE-like behavior of the scalar field before the recombination era can be achieved, but it is subjected to a high degree of approximations. One of the advantages of our approach is that one can find EDE-like behaviour of the scalar field more generically without any approximation and the result applies to a wide class of potentials.

\section{Conclusion}

In this study, we explore the dynamics and phase-space characteristics of a multi-component dark energy model. This model features a dark sector that includes a minimally coupled canonical scalar field and a cosmological constant. The equations are transformed into autonomous systems and subsequently into polar form. Since we are interested in making our analysis valid for a wide class of potentials, enabling us to study the multi-component models from a more general point of view, we have considered a parameterization for the potentials of the scalar field that helps us study a broad spectrum of potentials in a single setup.

By using dynamical system analysis, we find several fixed points that correspond to various cosmological eras together with possible late-time attractors. In this model, the late-time attractors are either completely dominated by the cosmological constant or present a scenario where both the scalar field and the cosmological constant coexist, excluding all other components.

We also investigated numerically the evolution of different cosmological parameters and compared it with the observed data. Most interestingly, our findings show the early time contribution of the scalar field occurring deep within the matter-dominated era, not near the recombination period. We also found that during this era of the scalar field, it behaves more like matter rather than dark energy, and the universe becomes more decelerated compared to the standard case. The total EOS of the universe also shows a transition from negative to positive values during this epoch. We acknowledge that a comprehensive parameter estimation for the choice of $\alpha$ parameters using a Boltzmann code like CLASS and current cosmological data is necessary to comment more accurately on this early-time contribution of the scalar field and how it could affect the overall evaluation and structure formation of the universe. Since this study focuses on the investigation of the phase space behaviour of the model, it is beyond the scope of the current work. It will be addressed in future research.

\begin{table*}[h]
    \centering
    \scalebox{1.3}{
    \begin{tabular}{|c|c|c|c|c|c|c|c|}
       \hline {Potentials} & {$N_{ini}$} & {$r_{ini}$} & {$\theta_{ini}$}  & {$\Omega_{m_{ini}}$} & {$\Omega_{r_{ini}}$} &  {$\lambda_{ini}$}    \\ \hline
         $V_{1}$ & $-10.1$ & $\sqrt{0.3\times 10^{-9}}$ & $tan^{-1}(-2)$ & $ 0.118 - \frac{(r_{ini})^{2}}{2} $ & $0.882-\frac{(r_{ini})^{2}}{2}$ & $-2.4$  \\  \hline
        $V_{2}$ & $-10.1$ & $\sqrt{0.3\times 10^{-9}}$ & $tan^{-1}(-2)$ & $ 0.118 - \frac{(r_{ini})^{2}}{2} $ & $0.882-\frac{(r_{ini})^{2}}{2}$ & $-2.93$  \\  \hline
        $V_{3}$ & $-10.1$ & $\sqrt{0.36\times 10^{-9}}$ & $tan^{-1}(-2)$ & $ 0.118 - \frac{(r_{ini})^{2}}{2} $ & $0.882-\frac{(r_{ini})^{2}}{2}$ & $-0.64$  \\  \hline
         $V_{4}$ & $-10.1$ & $\sqrt{0.3\times 10^{-9}}$ & $tan^{-1}(-2)$ & $ 0.118 - \frac{(r_{ini})^{2}}{2} $ & $0.882-\frac{(r_{ini})^{2}}{2}$ & $-4.49$  \\  \hline
          $V_{5}$ & $-10.1$ & $\sqrt{0.3\times 10^{-9}}$ & $tan^{-1}(-2)$ & $ 0.118 - \frac{(r_{ini})^{2}}{2} $ & $0.882-\frac{(r_{ini})^{2}}{2}$ & $-5.13$  \\  \hline
           $V_{6}$ & $-10.1$ & $\sqrt{0.3\times 10^{-9}}$ & $tan^{-1}(-2)$ & $ 0.118 - \frac{(r_{ini})^{2}}{2} $ & $0.882-\frac{(r_{ini})^{2}}{2}$ & $-5.785$  \\  \hline
           $V_{7}$ & $-10.1$ & $\sqrt{0.3\times 10^{-9}}$ & $tan^{-1}(-2)$ & $ 0.118 - \frac{(r_{ini})^{2}}{2} $ & $0.882-\frac{(r_{ini})^{2}}{2}$ & $-6$  \\  \hline
           $V_{8}$ & $-10.1$ & $\sqrt{0.3\times 10^{-9}}$ & $tan^{-1}(-2)$ & $ 0.118 - \frac{(r_{ini})^{2}}{2} $ & $0.882-\frac{(r_{ini})^{2}}{2}$ & $-6.59$  \\  \hline
         \end{tabular}
         }
    \caption{ The initial values of the variables for which the numerical solution of the system of equations Eq.$(\ref{polar_autonomous:1})$ has been obtained.}
    \label{tab:initiaconditions}
\end{table*}

\begin{table*}[!hbt]
    \centering
    \scalebox{1.3}{
    \begin{tabular}{|c|c|c|c|c|c|c|c|c|}
       \hline {Potentials} & {$\lambda_{ini}$} &  {$\alpha_{0}$}  & {$\alpha_{1}$} & {$\alpha_{2}$} &  {$z_{eq}$}  &  {$z_{\ast}$}  \\ \hline
         $V_{1}$ & $[-2.5, -2.4]$  & $[-20, -4.65]$ & $ [18 , 20] $ & $[0.4, 0.7]$ & $3228$ & $54$ \\  \hline
        $V_{2}$ & $[-3 , -2.93]$ & $0$ & $[16, 20]$ & $[0.9, 1]$ & $3228$ & $52$   \\  \hline
        $V_{3}$ & $[-0.7 , -0.5]$ &  $[-270 , -200]$ & $0$ & $[0.47, 0.8]$ & $3228$ & $58$ \\  \hline
        $V_{4}$ & $[-4.6 , -4.49]$  &   $[-8, -4.35]$ & $[2.8 , 3]$ & $0$ & $3293$ & $43$  \\  \hline
        $V_{5}$ & $[-5.3 , -5.13]$  & $[-16.5, -12]$ & $ 0 $ & $0$ & $3228$ & $44$ \\  \hline
        $V_{6}$ & $[-5.9 , -5.785]$ & $0$ & $[0.9, 1.1]$ & $0$ & $3228$ & $40$   \\  \hline
        $V_{7}$ & $[-6.1 , -5.99]$ &  $0$ & $0$ & $[-0.15,  -0.1]$ & $3228$ & $38$ \\  \hline
        $V_{8}$ & $[-6.9 ,- 6.59]$  &   $0$ & $0$ & $0$ & $3228$ & $31$   \\  \hline
         \end{tabular}
         }
    \caption{ The range of parameters and $\lambda_{ini}$ for which the qualitative behavior of the numerical solution of the system of equations Eq. $(\ref{polar_autonomous:1})$ remains unchanged.}
    \label{tab:parameterspace}
\end{table*}

\clearpage

\appendix

 \section{}\label{app:eigenvalues}

$$A_{\pm}=\frac{-\alpha _{1}\pm \sqrt{\alpha _{1}^2-4 \alpha _{0} \alpha _{2}}}{2 \alpha _{2}}.$$

$$E_{\mp}=\frac{-\alpha _{1}\pm \sqrt{\alpha _{1}^2-4 \alpha _{0} \alpha _{2}}}{2 \alpha _{0}},$$

$$ B_{\pm}=1+\frac{4\alpha_{2}}{\alpha_{0}}+\frac{4\alpha_{1}}{\alpha_{0}}E_{\pm}, $$

$$C_{\pm}=1+\frac{3\alpha_{2}}{\alpha_{0}}+\frac{3\alpha_{1}}{\alpha_{0}}E_{\pm},$$

$$D_{\mp}=\sqrt{-\frac{2}{\alpha_{2}}\left( \alpha_{1}A_{\mp} +\alpha_{0} \right)}.$$

\onecolumngrid
\vspace{3cm}
\section{ Observed data for the Hubble's parameter vs. redshift \& the e-folding.} \label{sec:data}

\begin{table*}[h]
    \begin{subtable}{0.5\linewidth}
        \centering
        \begin{tabular}{|c|c|c|c|}
        \hline
        $z$ & $N=\ln(\frac{1}{1+z})$ & $H(z)\left(\frac{\text{km/s}}{\text{Mpc}}\right)$ & Ref. \\
        \hline 0.07 &-0.067 & $69 \pm 19.6$ & \cite{Zhang_2014} \\
\hline 0.09&-0.086 & $69 \pm 12$ & \cite{PhysRevD.71.123001} \\
\hline 0.100 & -0.095& $69 \pm 12$ & \cite{PhysRevD.71.123001} \\
\hline 0.120 &-0.113 & $68.6 \pm 26.2$ & \cite{Zhang_2014} \\
\hline 0.170 & -0.157& $83 \pm 8$ & \cite{PhysRevD.71.123001} \\
\hline 0.179& -0.164 & $75 \pm 4$ & \cite{Moresco_2012}\\
\hline 0.199 &-0.181& $75 \pm 5$ &  \cite{Moresco_2012} \\
\hline 0.200 &-0.182& $72.9 \pm 29.6$ & \cite{Zhang_2014} \\
\hline 0.270 & -0.239& $77 \pm 14$ & \cite{PhysRevD.71.123001} \\
\hline 0.280 &-0.246& $88.8 \pm 36.6$ & \cite{Zhang_2014} \\
\hline 0.320 &-0.277& $79.2 \pm 5.6$ & \cite{Cuesta:2015mqa} \\
\hline 0.352 &-0.301& $83 \pm 14$ & \cite{Moresco_2012} \\
\hline 0.3802&-0.322 & $83 \pm 13.5$ & \cite{Moresco_2012} \\
\hline 0.400&-0.336 & $95 \pm 17$ & \cite{PhysRevD.71.123001} \\
\hline 0.4004&-0.336 & $77 \pm 10.2$ & \cite{Moresco_2012} \\
\hline 0.4247&-0.353 & $87.1 \pm 11.2$ & \cite{Moresco_2012} \\
\hline 0.440 &-0.364& $82.6 \pm 7.8$ &  \cite{Blake:2012pj}  \\
\hline 0.4497 &-0.371& $92.8 \pm 12.9$ & \cite{Moresco_2012} \\
\hline 0.470 & -0.385&$89 \pm 34$ & \cite{Ratsimbazafy:2017vga} \\
        \hline
        \end{tabular}
    \end{subtable}%
    \begin{subtable}{0.5\linewidth}
        \centering
        \begin{tabular}{|c|c|c|c|}
        \hline
        $z$ & $N=\ln(\frac{1}{1+z})$ & $H(z)\left(\frac{\text{km/s}}{\text{Mpc}}\right)$ & Ref. \\
        \hline 0.4783 &-0.390& $80.9 \pm 9$ &\cite{Moresco_2012} \\
\hline 0.480 &-0.392& $97 \pm 62$ & \cite{Stern:2009ep} \\
\hline 0.570&-0.451 & $100.3 \pm 3.7$ & \cite{Cuesta:2015mqa} \\
\hline 0.593 & -0.465&
$104 \pm 13$ & \cite{Moresco_2012} \\
\hline 0.600&-0.470 & $87.9 \pm 6.1$ &  \cite{Blake:2012pj}  \\
\hline 0.680&-0.518 & $92 \pm 8$ & \cite{Moresco_2012} \\
        \hline 0.730&-0.548 & $97.3 \pm 7$ & \cite{Blake:2012pj} \\
\hline 0.781 & -0.577& $105 \pm 12$ & \cite{Moresco_2012} \\
\hline 0.875&-0.628 & $125 \pm 17$ & \cite{Moresco_2012} \\
\hline 0.880&-0.631 & $90 \pm 40$ & \cite{Stern:2009ep} \\
\hline 0.900&-0.641 & $117 \pm 23$ & \cite{PhysRevD.71.123001} \\
\hline 1.037&-0.711 & $154 \pm 20$ & \cite{Moresco_2012} \\
\hline 1.300&-0.832 & $168 \pm 17$ &\cite{PhysRevD.71.123001} \\
\hline 1.363 &-0.859 & $160 \pm 33.6$ & \cite{Moresco:2015cya} \\
\hline 1.430&-0.887 & $177 \pm 18$ & \cite{PhysRevD.71.123001} \\
\hline 1.530&-0.928 & $140 \pm 14$ & \cite{PhysRevD.71.123001} \\
\hline 1.750&-1.011 & $202 \pm 40$ & \cite{PhysRevD.71.123001} \\
\hline 1.965&-1.086 & $186.5 \pm 50.4$ & \cite{Moresco:2015cya} \\
\hline 2.340&-1.205 & $222 \pm 7$ & \cite{BOSS:2014hwf} \\
\hline 2.360&-1.211 & $226 \pm 8$ & \cite{Blomqvist:2015pza} \\
\hline
        \end{tabular}
    \end{subtable}
    \caption{Observed data for $H(z)$ as a function of $z$ alongside the associated value of $N$, with references that have been used in this study. }
\end{table*}
\twocolumngrid
\clearpage
\bibliographystyle{unsrt}
\bibliography{sample}

\end{document}